\documentclass[epj,final]{svjour}

\usepackage{amsmath,amssymb}
\usepackage{bm}
\usepackage{color}
\usepackage{graphics}
\usepackage{amssymb}
\usepackage{amsmath}

\newcommand{\pa}{\partial}

\newcommand{\myref}[1]{(\ref{#1})}

\newcommand{\Om}{\Omega}
\newcommand{\de}{\delta}

\newcommand{\al}{\alpha}
\newcommand{\eps}{\varepsilon}

\newcommand{\Ga}{\Gamma}

\newcommand{\te}{\theta}
\newcommand{\Te}{\Theta}
\newcommand{\sig}{\sigma}
\newcommand{\sinc}{\text{sinc}}

\renewcommand{\leq}{\leqslant}
\renewcommand{\geq}{\geqslant}
\renewcommand{\tilde}[1]{\widetilde{#1}}
\renewcommand{\hat}[1]{\widehat{#1}}
\newcommand{\lan}{\langle}
\newcommand{\ran}{\rangle}
\newcommand{\nab}{\nabla}

\newcommand{\demi}{\frac{1}{2}}

\newcommand{\sur}[2]{{\displaystyle\mathop{#1}_{#2}}}

\newcommand{\mcal}[1]{\mathcal{#1}}

\newlength{\somme}
\newlength{\sommep}
\newlength{\sommebis}
\newlength{\sommepbis}
\begin{document}


\title{New conserved structural fields for supercooled liquids}
\authorrunning{J. Farago {\em et al.}}
\titlerunning{New fields for supercooled liquids}
\author{Jean Farago\inst{1} 
\and Alexander Semenov\inst{1}
\and Stefan Frey\inst{1}
\and J\"org Baschnagel\inst{1}}
\institute{Institut Charles Sadron, Universit\'e de Strasbourg, CNRS UPR 22, 23 rue du Loess--BP 84047, 
67034 Strasbourg Cedex 2, France}
\mail{jean.farago@ics-cnrs.unistra.fr}
\date{Draft version of \today}

\abstract{
By considering Voronoi tessellations of the configurations of a fluid, we propose two new conserved fields, which provide  structural information not fully accounted for by the usual 2-point density field fluctuations (structure factor). One of these fields is scalar and associated to the Voronoi cell volumes, whereas the other one, termed the ``geometrical polarisation'',  is vectorial, related to the very local anisotropy of the configurations. We study the static and dynamical properties of these fields in the supercooled regime of a model glass-forming liquid. We show in particular that the geometrical polarisation is both statically correlated to the force field and contrary to it develops a plateau regime when the temperature is lowered. We attribute this behaviour to the microsopic disorder of the underlying inherent structures (IS) which dictate the dynamics on time scales larger than the true microscopic time, in the strong supercooled regime. In this respect, this work raises the issue of to what extent the inter IS dynamics, intrinsically anisotropic and collective (cf. T.B. Schr\o der et al. {\it J. of Chem. Phys.}, {\bf 112}, 9834 (2000)), could be related to their polarisation field. 
}


\maketitle

\section{Introduction}

Among the different theoretical {\em predictive} approaches to the physics of glass-forming liquids, the mode-couplig theory (MCT) \cite{KobBinder,Goetze} is among the few  descriptions rooted at a true microscopic level (see also \cite{ParisiRMP,Schweizer2005JCP}). Indeed, the starting point of MCT is the set of  evolution equations for the microscopic conserved fields, wherefrom all physical predictions are derived. However, this sturdy basis does not prevent the MCT from shortcomings of its own, due to approximations necessary to get tractable equations.
 First, it is well-known that the MCT predictions are not fully satisfactory, insofar as a spurious ergodic transition at some temperature $T_c$ is predicted by the theory and never observed in simulations or experiments \cite{KobBinder}. Besides, in the temperature range where the MCT works at best, the fits with experiments/simulations are obtained at the price of a temperature rescaling, whose origin remains elusive. Secondly, the MCT results are somewhat difficult to interpret physically, owing to the fact that the geometrical coordinates are encoded by their Fourier modes. This point becomes even more problematic for temperatures close to $T_c$, for it has been hinted that the diffusion process becomes progressively more local (and activated) when the temperature is lowered \cite{BerthierBiroliRMP,BerthierBook,SchweizerCurrOp2007}. In a Fourier perspective, this locality is translated  by  finely adjusted relative amplitudes and phases of different modes, a tailoring which is presumably lost by the approximations inherent to the MCT approach.

The analysis of the limitations of MCT (and its connections with spin glass systems) has elicited other approaches which shed new light on the physics of glass-forming liquids: phase space studies, facilitation models, new concepts (point-to-set length, propensity,\ldots), random first-order transition, etc\ldots (see the recent reviews \cite{BaschnagelVarnik,CavagnaReport,BerthierBiroliRMP}). Amidst the flourishing of these alternative approaches, the MCT appears today to some extent isolated (with the noteworthy exceptions of \cite{InhomMCT_PRL} and \cite{ZFGST}). A theoretical framework, able to reconcile on the one hand the modal MCT approach, certainly grasping the physics of moderately supercooled liquids and on the other hand  real-space features described by facilitation models, preponderant in strongly supercooled systems,  is both lacking and desirable.

A possible path toward this synthesis would be to extend the MCT beyond the ideal theory. Attempts in this direction have already been undertaken early \cite{GoetzeSjoegrenEMCT,DasMazenkoEMCT}; the avoiding of the MCT transition had been sought  considering the momentum field relaxation as a supplementary relaxation channel \cite{KobBinder}. However, these approaches have been  seriously challenged in \cite{Catesramaswamy,LiuOppenheim}. The underlying philosophy of these extensions is to recognize that a conserved field could act as a supplementary decay channel (beyond the density fields), through which the system would relax preferentially when the temperature is lowered. This general idea is certainly valuable, although the missing relevant decay channels are probably not only  related to the momentum field (Ref. \cite{Catesramaswamy} stressed the fact that this field is irrelevant for glass-forming colloids, where the ergodic transition is nevertheless avoided as well \cite{SchweizerCurrOp2007}). Were it be possible to put forward some {\em other} conserved fields, a new route toward an extension of the ideal MCT theory would open.

This paper intends to show that  it is possible to devise some new conserved fields, which would be {\em not  redundant with the density field in any approximate fluid theory}. From the strict physical point of view, it is obvious that the sole conserved field of, say,  a monodisperse colloidal system (for sake of simplicity, we disregard the fact that this system would crystallize) is its density field $\rho(\bm k)=\sum_j \exp(i\bm k\cdot\bm r_j)$ (expressed in Fourier modes). As a result, one cannot enlarge the set of conserved quantities by adding a field which would be conserved by the virtue of a physical property distinct from the mass conservation. For instance, any field depending solely on the microscopic configuration (and not on the velocities) can be related to $\rho(\bm k)$. However, as soon as the knowledge of the density field is not total, due e.g. to an implicit coarse-graining, or a truncation in a hierarchy of dynamical equations, this one-to-one relation between density field and functions over the configurational phase space ceases to be true.
In particular, we will show that thanks to the Voronoi tessellations,  new conserved configurational fields naturally emerge. The Voronoi tessellation is a mathematical partition of the physical space and introduces  the notion of neighbourhood of a particle, which is both very intuitive and related to the density field by an extremely complicated functional dependence. The outcome of this intricacy is precisely to allow a view on the structural properties which is only partially accounted for by the traditional moments of the density distribution.

In a first part (section \ref{secvol}), a so-called {\em volume field} is  defined for each fluid configuration using the Voronoi cells of the configuration. Some fundamental static and dynamical properties of this notion are presented, from which a second conserved field, vectorial and Voronoi based, is deduced. This vectorial field, termed {\em geometrical polarisation}, is investigated in section \ref{geopol}. It is shown that this field naturally couples to the force field, but contrary to it develops a plateau relaxation when the temperature is lowered. This paper should be considered as an introductory work to these  volume and polarisation fields, which are sensible candidates for devising a new type of extended MCT (eMCT). We do not aim here to write down and solve this eMCT, a task we postpone for future publication. We just stress that this program, although difficult, is not impossible, insofar as the static correlators needed in the eMCT have explicit expressions in terms of geometrical features of the Voronoi cells.

Our analysis employs equilibrium trajectories from molecular dynamics simulations of a bead-spring model for glass-forming polymer melts. The model and the simulations are explained in Ref. \cite{stefanthese,stefanMCTpaper}. Here we only give a brief summary. We examine a melt, made of 3072  oligomer chains of 4 monomers of mass $1$. The nearest neighbour intra-chain monomers interact via a  potential $V_{\rm intra}(r)=\demi k(r-\ell_0)^2$, with $k=1110$ and $\ell_0=0.967$. The other pairs of monomers, including  not nearest neighbour intra-chain monomers, interact via a Lennard-Jones potential with $\eps=1$, $\sig=1$,  cut off and shifted at $r_c=2.3$. The dynamics is thermalized  with a  Nos\'e-Hoover thermostat at constant volume. The volume for different temperatures is determined so as to maintain a quasi-zero constant pressure . For this model,  $T_c\simeq 0.38$ has been determined by usual techniques \cite{stefanthese,stefanMCTpaper,KobBinder}.

\section{The volume fields and the generalized structure factors\label{secvol}}

\subsection{Definitions, Basic properties\label{DBP}}

For a fluid of $N$ particles whose positions are $\bm r_j$ ($j=1\ldots N$), we denote by $U_j$ the volume of the Voronoi cell around the particle $j$ and by $v$ its average. This volume is defined as the set of points closer to $j$ than to any other particle \cite{SpatialTessellations}. The Voronoi cells are convex polyhedrons enclosing the particle $j$.

  The Voronoi tessellations (i.e. the partition of space defined by the set of Voronoi cells for a given configuration) have been studied many times in the context of liquids \cite{Rahman,Glotzer_Voronoi,KumarKumaran}: In \cite{Glotzer_Voronoi} for instance, the distribution of volumes and asphericities of Voronoi cells are computed for glass-forming liquids, and display noticeable universal features as well as fluctuations scaling $\propto T^{1/4}$ with respect to the temperature at constant volume.

 In dense liquids one expects the fluctuations of $U_j$ to be small as a result of the weak compressibility of the system: A substantial fluctuation of $U_j$, say positive, is likely to occur by shoving the surrounding particles away from the $j$-th particle. But, as noticed in \cite{Glotzer_Voronoi}, the compressibility is not easily related to $\sig_v^2$, the standard deviation of $U_j$; One does not observe $\chi_T\sim \sig_v^2/[vk_BT]$. The reason for that is that during an increase of $U_j$, the test particle is assumed to remain within the inflating volume, whereas $\chi_T$ is related to the fluctuations of the number of particles within a fixed volume : This slight difference does  matter for such a small system (for large systems, fluctuations of particles at constant volume and the reverse are related by a density term, a density which precisely is a strongly fluctuating observable at the level of the particle).
Actually, the fluctuations of $U_j$ have two distinct origins: on the one hand, different (i.e. not superimposable) configurations of the first shell of neighbours give rise to fluctuations of $U_j$. On the other hand, different positions of the tagged particle within its cell induce different values of $U_j$ and hence, fluctuations. The first contribution is  linked to the compressibility, whereas the second is not. Although mathematically well defined, this distinction is indeed somewhat artificial, because such a partitioning does not correspond dynamically to well separated timescales. 

\begin{figure}
  \centering \resizebox{7cm}{!}{\includegraphics{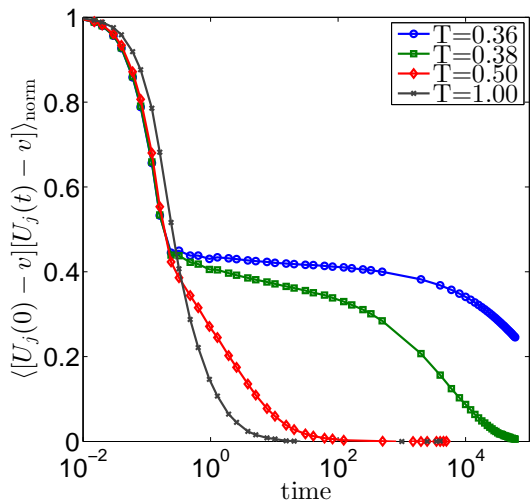}}
  \caption{Normalized autocorrelation of $U_j$ at different temperatures (note that $T_c\sim0.38$).}
  \label{autocorr_UU}
\end{figure}

In fig. \ref{autocorr_UU}, we
plot the normalized autocorrelation $\lan [U_j(0)-v][U_j(t)-v]\ran/\lan [U_j-v]^2\ran$ at several temperatures. The first rapid decay is related to the fast vibrational motions of all particles with respect to their local metastable equilibrium positions. The subsequent  plateau, which develops for low temperatures, is the embodiment  of the transient arrest of the dynamics of the density field expressed within the variables $U_j$. This is an obvious statement, but there is however something interesting associated to it: The initial decay of the excess/default of surrounding volume is  unsufficient to smooth out completely the cell volume fluctuations from particle to particle (otherwise the curves would have nearly vanished), therefore the crossover to a nonzero plateau should account for the structural disorder of the underlying  inherent structure (IS). The presence of the positive plateau tells us that this IS is  correlated at the level of the local structure to the initial condition it comes from: An initial condition with a positive (resp. negative) fluctuation of $U_j$ is likely to correspond to an IS with also a positive (resp. negative) fluctuation of $U_j$. If one focuses now on the dynamics on a larger timescale (``large'' means here for timescales comparable to the beginning of the plateau in Fig. \ref{autocorr_UU}), when it can be envisioned as a hopping from IS to IS \cite{Schroderetal}, an interesting question would be to determine to what extent the residual Voronoi volume field (i.e. the nonzero field associated to the IS) determines the transition to the next IS. For instance, if for an IS $U_j>v$, the choice by the dynamics of the next IS should be biased so as to most probably lessen  the value of $U_j$ (for sake of simplicity, we assume that the median value of $U_j$ in the IS is $v$). This example is certainly a little bit too simple, for if the Voronoi volumes  happen to be significant in the inter IS dynamics, this would  involve necessarily their distribution over larger regions than just one Voronoi volume, owing to the fact that the transition from one IS to the next typically involves a dozen of particles \cite{Schroderetal}.

\subsection{Volume field}

The  Voronoi volumes are obviously not conserved individually, but globally due to the partitioning of space  by the Voronoi tessellation. We are thus led to define, for a fluid of $N$ particles whose positions are $\bm r_j$ ($j=1\ldots N$),  the volume field by
\begin{align}
  \rho_v(\bm r,t)&=v^{-1}\sum_{j=1}^NU_j(t)\de(\bm r(t)-\bm r_j(t))\label{rhov}
\end{align}
 This field has the significant property of being a conserved field, in the sense that the quantity $U_j$ attached to the particle $j$  varies  by exchange with the adjacent  Voronoi volumes. As already noted,  this field is in principle entirely known as soon as the configuration of the particles is known (and vice-versa), but this subordination is actually only formal.

\medskip

From this field we can define three different generalized structure factors. The first two are
\begin{align}
  S_v(k)&=N^{-1}\lan\rho_v(-\bm k)\rho_v(\bm k)\ran\ \ \text{(VSF)}\\
S_i(k)&=N^{-1}\lan\rho(-\bm k)\rho_v(\bm k)\ran\ \ \text{(ISF)}\\
\text{with }\rho_v(\bm k)&\equiv v^{-1}\sum_{j=1}^NU_j\exp(i\bm k\cdot \bm r_j)
\end{align}
Notice that $\rho_v(\bm k,t)$ is the Fourier transform of $\rho_v(\bm r,t)$  and we will omit the time henceforth in equal time correlation functions, for sake of clarity. The volume structure factor (VSF) $S_v(k)$ is the exact counterpart for $\rho_v$ of the usual structure factor (SF) $S(k)=N^{-1}\lan\rho(-\bm k)\rho(\bm k)\ran$. The intermixed structure factor (ISF) $S_i(k)$ is the cross-correlation of the fields $\rho(\bm k)$ and $\rho_v(\bm k)$, and is likely to be nonzero since these two fields share exactly the same symmetries.

A third generalized structure factor can be constructed by orthogonalizing $\rho_v(\bm k)$ and  $\rho(\bm k)$ with respect to the canonical average:
\begin{align}
\rho_{v\perp}(\bm k)&\equiv \rho_v(\bm k)-\frac{S_i(k)}{S(k)}\rho(\bm k)\\
  S_{v\perp}(k)&=N^{-1}\lan \rho_{v\perp}(-\bm k)\rho_{v\perp}(\bm k)\ran\ \ \text{(OVSF)}\nonumber\\&=S_v(k)-S_i(k)^2/S(k)\label{svperpdef}
\end{align}
where OVSF stands for orthogonalized volume structure factor. $\rho_{v\perp}(\bm k)$ is defined so as to get $\lan\rho_{v\perp}(-\bm k)\rho(\bm k)\ran=0$ for all $k$.

\bigskip

\begin{figure}
\centering\resizebox{7cm}{!}{\includegraphics{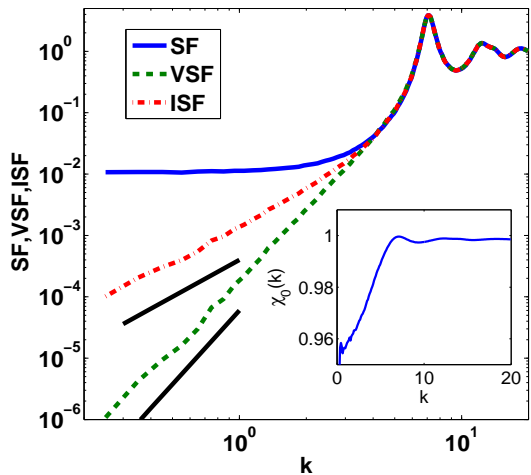}}
\caption{(color online) Log-log plot of $S(k)$ (blue solid), $S_v(k)$ (green dashed) and $S_i(k)$ (red dash-dotted). The solid thick black lines highlight the slopes 4 and 2 (respectively, from bottom to top). The inset represents the correlation coefficient of $\rho(\bm k)$ and $\rho_v(\bm k)$.\label{fig1_starter}}
\end{figure}

The typical behaviour of the SF, VSF and ISF for our oligomer melt  is plotted in figure \ref{fig1_starter}.  Two distinct behaviours emerge according to the value of $k$. 
For small $k$ spanning from  the hydrodynamic range (i.e. the $k\rightarrow 0$ domain where the SF has a well-defined constant plateau value) to the crossover domain (i.e. up to $k\sim 4$), the SF, VSF and ISF behave very differently: Whereas the SF reaches for decreasing values of $k$ a nonzero compressibility plateau, the VSF and ISF both vanish, the former $\propto k^4$ and the latter $\propto k^2$. The fact that all volume structure factors vanish in the hydrodynamic limit is not surprising; It corresponds to the (somewhat imprecise) common sense statement according to which the volume cannot fluctuate (this point will be developed further in the next subsection). What is more dumbfounding is the different exponent associated to this asymptotic behaviour. This nontrivial point will be analysed in section \ref{geopol}.

\begin{figure}
\centering  \resizebox{7cm}{!}{\includegraphics{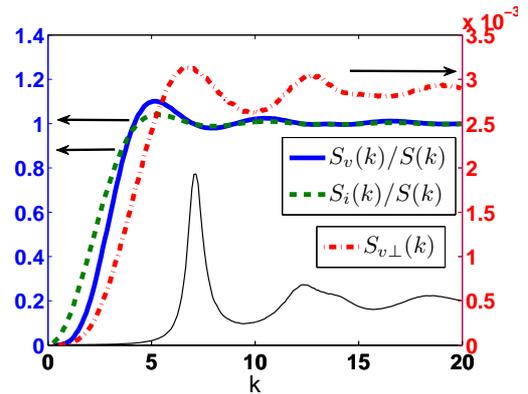}}
  \caption{(color online) Typical shape of $S_v(k)/S(k)$ (solid, left scale), $S_i(k)/S(k)$ (dashed,left scale) and the OVSF $S_{v\perp}(k)$ (dash-dotted, right scale). The thin solid black graph in the bottom shows the SF with arbitrary units for sake of comparison.}
  \label{fig2_Svperp}
\end{figure}

The second main region in  $k$ space concerns the microscopic domain, i.e. for $k\gtrsim 4$. Fig. \ref{fig1_starter} indicates that the three SFs are here strongly alike. The solid and dashed curves in Fig. \ref{fig2_Svperp}, showing $S_v(k)/S(k)$ and $S_i(k)/S(k)$ respectively,  make this point more precise: One observes that a weak structuration (oscillations) of the signal around 1, out of phase with that of the SF itself (black thin curve, arbitrary ordinate units). This shows that $\rho_v$ and $\rho$ are quite strongly correlated in the microscopic domain, a fact that can be quantified using the correlation coefficient $\chi_0(k)=S_i(k)/\sqrt{S_v(k)S(k)}$. The Cauchy-Schwarz inequality imposes that $-1\leq \chi_0(k)\leq 1$, a value of 1 meaning an exact positive proportionality of $\rho$ and $\rho_v$. The correlation coefficient $\chi_0(k)$ is shown in the inset of Fig. \ref{fig1_starter}, where
the strong correlation between the two variables is confirmed over the whole $k$ range. Roughly speaking, a large part of information carried by $\rho_v$ is already provided by $\rho$, {\em at the level of the second-order fluctuations}. However, a small share of $\rho_v$ provides information of its own, not accounted for by the SF. The orthogonalized version of $\rho_v$, namely $\rho_{v\perp}$, precisely aims at ``freeing'' the volume density field from all the redundancy of statistical information carried both in $\rho$ and $\rho_v$. As a result, the typical values of the associated structure factor  $S_{v\perp}(k)$ (red dash-dotted in Fig. \ref{fig2_Svperp}) are quite small. The OSVF starts from zero for $k\rightarrow 0$, grows $\propto k^4$ (this is easily seen from \myref{svperpdef}), and reaches a plateau, weakly modulated in phase with the SF, with typical values $\sim 3\times 10^{-3}$. The asymptotic plateau value is nothing but the normalized variance of the Voronoi volume $\mcal{V}_n=[\lan U_j^2\ran-\lan U_j\ran^2]/\lan U_j\ran^2$, therefore the typical small values of $S_{v\perp}(k)$ (or equivalently, $1-\chi_0(k)$) mirror the smallness of the available volume fluctuations  around the particles in the dense phases (notice  that a typical value of the available volume fluctuation is $\sim \sqrt{\mcal{V}_n}\sim 5.4\%$ of the mean Voronoi volume). For sake of comparison, the ideal gas displays a variance $\mcal{V}_n\sim 0.18=(0.42)^2$ \cite{gilbert}.

A natural reflex would be to dismiss the field $\rho_{v\perp}$ (and therefore $\rho_v$ as well), as being unable to provide any valuable physical input, because of the smallness of its typical fluctuations (with respect to $S(k)$ for instance). Such a reflex should be at least deferred if one gives some credence to the mode-coupling theory, because  a key criterion for the relevance of a variable as a decay channel is  its ability  to couple to the generalized force expression within the memory kernel. The nonlinear character of the theory is moreover capable of inducing strong dynamical amplifications of tiny modifications of the control parameters (the strong slowing down of the dynamics for only a small change of temperature for instance).
Furthermore,  MCT deals actually only with  normalized variables \cite{kawasaki_MCT_ovdpd}, therefore their actual unnormalized level of fluctuations does not play any direct role.

 The next paragraph  gives a hint at the physical content of this field by considering a specific example.

\subsection{Physical content: An illustration}

\begin{figure}
\centering  \resizebox{7cm}{!}{\includegraphics{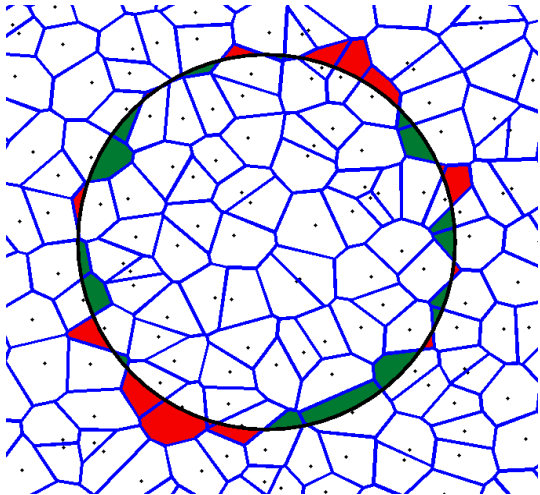}}
  \caption{(color online) The field $\rho_v(\bm r)-\lan\rho_v\ran$, integrated over the disc, measures the mismatch between the area covered by the Voronoi cells of inside particles and the disc area. This integral is the difference between the outside (red)  and   inside (green) shaded areas.}  \label{ex_circle}
\end{figure}

Let us consider the virtual spherical volume $V$ (radius $R$) with its center located at $\bm r=0$ (arbitrary origin), drawn into a fluid at equilibrium. We define  the following observable:
\begin{align}
\mcal{I}_v&=  \int_V [\rho_v(\bm r)-v^{-1}]d^3r=v^{-1}[V_v-V]
\end{align}
which is the fluctuation of $\rho_v(\bm r)$ with respect to its mean value $v^{-1}$, integrated over the sphere. $V_v$ is the volume obtained by the aggregation of all Voronoi cells located inside $V=4\pi R^3/3$. Therefore, the integral $\mcal{I}_v$ is essentially a {\em boundary term}, as exemplified  in figure \ref{ex_circle} (the picture is in 2D, and for an ideal gas configuration, for sake of illustration, but we assume a 3D fluid in the discussion). The value of $\mcal{I}_v$ can be read by adding the external shaded region and substracting the internal ones. 
By contrast, a similar integral $\mcal{I}$, where the density field replaces the volume field is  the fluctuation of the particle number inside the sphere, and therefore a bulk term. As a result, the statistical signature of the fluctuations of $\rho_v$ are likely to account for a physics somewhat more local than that of $\rho$. Let us also notice the nonconnected topology of $\mathbb{R}^3\setminus [V_v-V]$, which implies for instance that a diffusing particle initially at $r<R$ cannot diffuse on distances larger than $R$ without coupling to $\mcal{I}_v$. Another example is the following conditionality: As long as $\mcal{I}_v$ does not relax, it prevents the homogeneisation of the inside and the outside and therefore would sustain an heterogeneity, if the initial densities in the two regions were initially different.

\medskip

As $\mcal{I}_v$ is actually a surface term, one expects its fluctuations to be affected by this peculiar dimensionality. Indeed, one has
\begin{align}
  \lan\mcal{I}_v^2\ran&\sur{\sim}{R\rightarrow\infty} 4R^2v^{-1}\int_0^\infty dk k^{-2}S_v(k)\label{IvIv}\\
\lan \mcal{I}_v\mcal{I}\ran&\sur{\sim}{R\rightarrow\infty} 4R^2v^{-1}\int_0^\infty dk k^{-2}S_i(k)\label{IvI}
\end{align}
To demonstrate this, we begin by noticing that we have omitted so far the singular part that all generalized structure factors have in common with the ordinary SF: For instance, one has $S_v(k)=S_{v,\rm reg}(k)+(2\pi)^3v^{-1}\de(\bm k)$. Implicitly, the SF cited in \myref{IvIv} and \myref{IvI} are the regular parts. This implicit will be continued. Thanks to the singular part, we arrive to $\lan \mcal{I}_v^2\ran=\int (d^3k/(2\pi)^3) S_v(k)|\Pi(k)|^2$ where $\Pi(k)=4\pi k^{-3}[\sin(kR)-kR\cos(kR)]$ is the Fourier transform of the indicator function of the sphere. Owing to the fact that $S_v(0)=S_i(0)=0$, the main term in the limit of large $R$ is given by the last term of $\Pi(k)$, whence the result (after the neglecting of the rapid oscillating part of $\cos^2(kR)$).

On a qualitative level, the $R^2$ dependence in \myref{IvIv} is explained by considering the various Voronoi cells mismatches as $\sim R^2$ independent random variables with zero mean. The average of their sum squared is thus $\sim R^2$ as well. 

For the result \myref{IvI}, the reasoning is  different: A, say, positive fluctuation of density inside the sphere makes the Voronoi cells slightly smaller than those outside. This induces a typical  {\em positive} mismatch of the boundary Voronoi cells: The inner particles at the boundary, denser,  are typically closer to the sphere frontier than the outer particles. As a result, the frontiers of the Voronoi cells of inside boundary particles tend to overstep the limits of the sphere, since, by construction, a facet  between two Voronoi cells is at equal distance from the two particles concerned.
 This typical mismatch is in the linear order proportional to $[N_v-\lan N_v\ran]/\lan N_v\ran$, whence $\lan \mcal{I}_v\mcal{I}\ran$ is proportional to $R^2\times \lan[N_v-\lan N_v\ran]^2\ran/\lan N_v\ran\sim R^2$ as before.

This  example makes more precise the intuitive statement according to which the volume should not fluctuate at large lengthscales: We see that indeed the large scale bulk fluctuations are exactly compensated, the remaining ones being squeezed into a surface region not thicker than the typical Voronoi cell size. This exact compensation endows $\mcal{I}_v$ with fluctuating properties very different from that of $\mcal{I}$, and accounted for in the large $R$ limit only if $\lim_{k\rightarrow 0} S_{v,i}(k)=0$. It must however be noted that this example does not enforce any particular functional form for $S_{v}$ and $S_i$ near $k=0$ provided they goes to zero. For symmetry reasons, they are at least $\propto k^2$, but Eq. \myref{IvIv} does not hint at why $S_v(k)$ should be $\propto k^4$ in the hydrodynamic limit. This nontrivial behaviour will be analysed in section \ref{geopol}.

\section{Geometrical polarisation} \label{geopol}

What is remarkable in the low $k$ region is the fact that if $S_i(k)$ goes to zero as $k^2$, $S_v(k)$ goes to zero as $k^4$, a fact that is not obvious from e.g. the inspection of  formulas \myref{IvIv} and \myref{IvI}. One just notices that $S_v(k)\propto k^4$ removes  long wavelength contributions  from the integral in \myref{IvIv}, which is qualitatively in accordance with the physical content of $\mcal{I}_v$.

\begin{figure}
\centering  \resizebox{7cm}{!}{\includegraphics{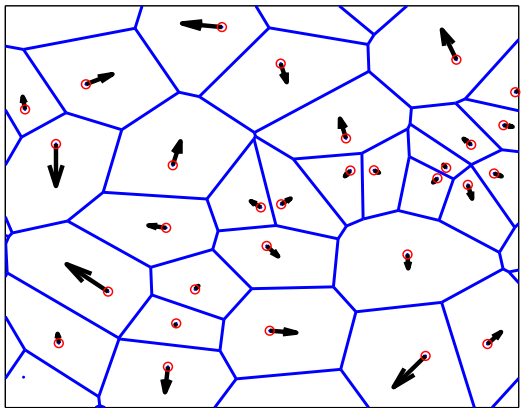}}
  \caption{(Color online) An example of the geometrical polarisation field for a 2D Poisson-Voronoi (ideal gas) diagram. The arrows represent the vectors $\bm\tau_j$, have their origins at the particle's locations (red circles). Note that their tip is not at the geometrical centers (centroid) of the Voronoi cells the particles are associated to, because of the  factor $U_j/v$ in the definition of $\bm\tau_j$. }\label{sketch_geopol}  
\end{figure}

Actually, this property is related to a remarkable and general feature of the Voronoi tessellations, which, to the best of our knowledge, had not been unveiled so far. We will prove in Appendix \ref{App_pol_2} that the total geometrical polarisation of the configuration, defined by 
\begin{align}
  \bm P&=v^{-1}\sum_iU_i[\bm s_i-\bm r_i] \label{Pdef}
\end{align}
where $\bm s_j$ is the barycenter of the $j$-th Voronoi cell, is independent of the configuration (in a set of configurations connected by continuous transformations). More precisely, this is strictly true only for a system without boundaries (infinite or with a toroidal geometry), for which one has moreover $\bm P=\bm 0$. For systems with boundaries $\bm P$ is constant within subsets of configurations with identical boundary Voronoi cells, and one has just a local conservation in the bulk. We also define  the microscopic {\em geometrical polarisation field} by
\begin{align}
  \bm p(\bm r)&=\sum_j \bm \tau_j\de(\bm r-\bm r_j)\\
\bm \tau_j&\equiv v^{-1}U_j(\bm s_j-\bm r_j)\label{taudef} 
\end{align}
A typical realization of this field is shown in Fig. \ref{sketch_geopol} for a 2D Voronoi diagram (for sake of clarity) corresponding to a random choice of points with a density 1. Note that a clear anticorrelation of the polarisations arises when two particles are close to each other. This reflects the symmetrical position of two adjacent particles with respect to their common dividing facet.

Two distinct propositions have to be demonstrated. First, assuming that $\bm P$ is indeed conserved and zero for an infinite system, we show that this implies a $k^4$ behaviour of $S_v(k)$ (and subsequently, of $S_{v\perp}(k)$). Second, we show that $\bm P$ is indeed conserved. These two important but technical demonstrations are developed in the Appendices \ref{App_pol_1} and \ref{App_pol_2}, respectively.

\subsection{Static properties of the geometrical polarisation}

\begin{figure}
\centering  \resizebox{7cm}{!}{\includegraphics{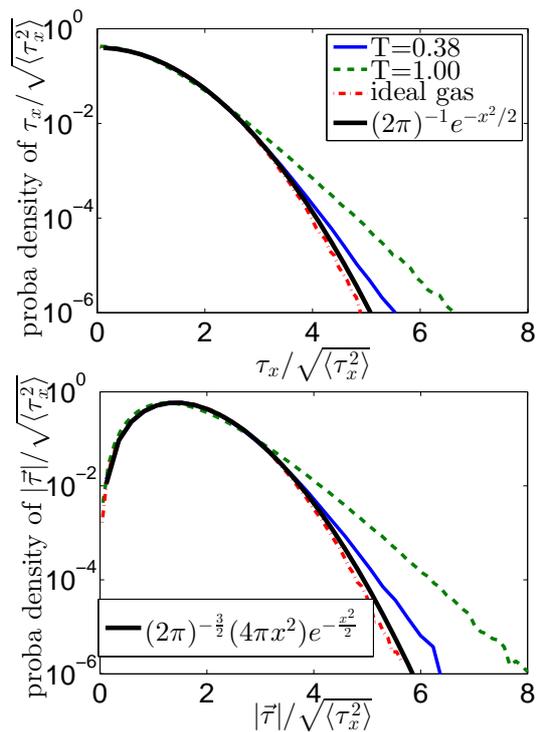}}
  \caption{(color online) probability densities of the scaled variables $\tau_x/\sqrt{\lan \tau_x^2\ran}$ (left) and $|\bm\tau|/\sqrt{\lan\tau_x^2\ran}$ (right). The $T=0.38$ and $T=1.00$ curves refer to the system defined in the introduction, and the formulas are those of the Maxwell-Boltzmann theory. }  \label{pol_fluct}
\end{figure}

The simplest characterization of $\bm\tau_j$ is probably its equilibrium distribution. In fig. \ref{pol_fluct} are shown the distributions of a $\tau_{j,x}$, an arbitrary cartesian component of the polarisation (upper plot) and that of its modulus $|\bm\tau_j|$ (lower plot).
These distributions are  well described by the  Maxwell-Boltzmann (MB) distributions for the velocities components and modulus in a gas, except for the high values of $\tau_x$ (or $|\bm\tau|$), where the simulated distributions are larger than the MB distributions. By conservation of probability, this implies a tiny depletion in the low $\tau$ region, invisible in the semilog representation. This departure from the MB distribution is quasi-absent for the ideal gas, and tends to be reduced for our system when lowering the temperature.

To understand why the distributions of $\tau_x$ and $|\bm\tau|$ comply  with the MB distribution, we remark that (in 2 or 3 dimensions)
\begin{align}
  \bm\tau_j&=\frac{1}{2v(d+1)}\sum_{i\big/\lan i,j\ran}r_{ji}S_{ji}\bm s_{ji}\label{centroidformula}
\end{align}
where the sum is over the particles $i$ sharing a facet with $j$, $r_{ji}=|\bm r_i-\bm r_j|$, $S_{ji}$ is the area of the common facet, and $\bm s_{ji}$ is the vector joining the particle $j$ to the barycenter of the facet $S_{ji}$. This formula comes from the fact that the $i$-th Voronoi cell is the stacking of triangles (2D) or tetrahedra (3D) whose the $S_{ij}$ are the basis and the particle $i$  the common summit \cite{SpatialTessellations}; Each term of the sum is just the volume of the simplex times the vector joining $i$ to the simplex' centroid. 

Eq. \myref{centroidformula} shows that in 3 dimensions, $\bm\tau_j$ is typically the sum of $\sim 14-15$ terms which are only weakly correlated to each other, for it is well-known that the structural correlations in a dense fluid does not extend significantly beyond the first shell of neighbours. As a result $\tau_{j,x}$ resorts approximately to the central limit theorem, and must have an approximate Gaussian distribution. The isotropy of the direction of $\bm\tau$ leads also to the MB distribution for the distribution of moduli.

The observed departures from the MB distribution for our system (solid and dashed lines in fig. \ref{pol_fluct}) come from the polymeric nature of the fluid, and in particular from the bond length between two adjacent monomers in a chain. This bond length is quite rigid, and slightly shorter than the typical distance between two neighbouring particles. This situation is likely to deplete slightly the occurence of very small polarisation, and conversely gives rise  to extra high-valued polarisations.


The magnitude of the polarisation with respect to the temperature is indicated in table \ref{table}.
\begin{table}
\begin{center}
\begin{tabular}{|c|c|c|}
\hline
 Temperature & $\de\tau$ &$\displaystyle\chi_{F\tau}$\\
\hline
$0.38$ & 0.0374 &0.3577\\
$0.50$ & 0.0438 &0.3338\\
$1.00$ & 0.0751 & 0.2190\\
ideal gas &  0.2141 &\\
\hline  
\end{tabular}
\caption{$\de\tau\equiv\sqrt{\lan\bm\tau^2\ran}/v^{1/3}$ is the standard deviation of polarisation, normalized by $v^{1/3}$. The third column is $\chi_{F\tau}=\lan \bm F_j\cdot\bm \tau_j\ran/[\lan \bm F_j^2\ran\lan \bm \tau_j^2\ran]^{1/2}$, the correlation coefficient between the total force on a monomer and its polarisation.}
\end{center}
\label{table}
\end{table}
One observes that the polarisation does not exceed few percents of the typical nearest-neighbour distance, and that it decreases with decreasing temperature. It is worth noticing that the $T\rightarrow 0$ limit of the normalized  standard deviation of the polarisation is probably not zero (we neglect  here the possibility of a thermodynamic transition) because such a case would correspond asymptotically to centroidal Voronoi tessellations, i.e. configurations such that each particle occupies the barycenter of its Voronoi cell. We do not see which could be the thermodynamic force responsible for such an arrangement (within configurations which would be anyway disordered; Voronoi centroidal {\em and} disordered configurations do exist). On the contrary, one expects for the standard deviation of the polarisation a $T\rightarrow0$ limit corresponding to the (common) value of any inherent structure.

\medskip

\begin{figure}
\centering \resizebox{7cm}{!}{\includegraphics{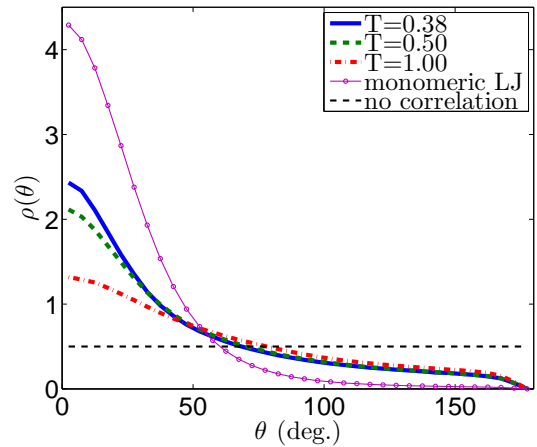}}  
  \caption{Probability density function $\rho(\theta)=dP(\te)/[\sin\te d\te]$ of the angle between the polarisation $\bm\tau_j$ and the total force $\bm F_j$ acting on monomer $j$. The plots associated to a temperature refer to the oligomer melt; the ``monomeric LJ'' plot (line+circles) refer to a metastable Lennard-Jones fluid at density $1.0625$ and $T=1.0$. For a non-correlated vector pair, one would have the constant line $\rho(\te)=1/2$.}
  \label{corr_angle_force_tau}
\end{figure}

Provided one considers only static correlations, a consequent correlation between the total force $\bm F_j$ applied on the monomer $j$ and the polarisation $\bm\tau_j$ does exist, see table \ref{table}, and fig. \ref{corr_angle_force_tau}.
 This is simply related to the fact that if one moves a particle toward another one, the total force and the polarisation react the same way, antiparallel to the direction of motion. Therefore, one is tempted to envision $\bm\tau$ merely as a  sophisticated ``ghost'' of the force, and thus useless. We will see that this is not true, because both vectors decorrelate very differently with time: There is more in the geometrical polarisation than just a vector correlated to the force.

\subsection{Autocorrelation of $\bm\tau_j$}

\begin{figure}
\centering  \resizebox{7cm}{!}{\includegraphics{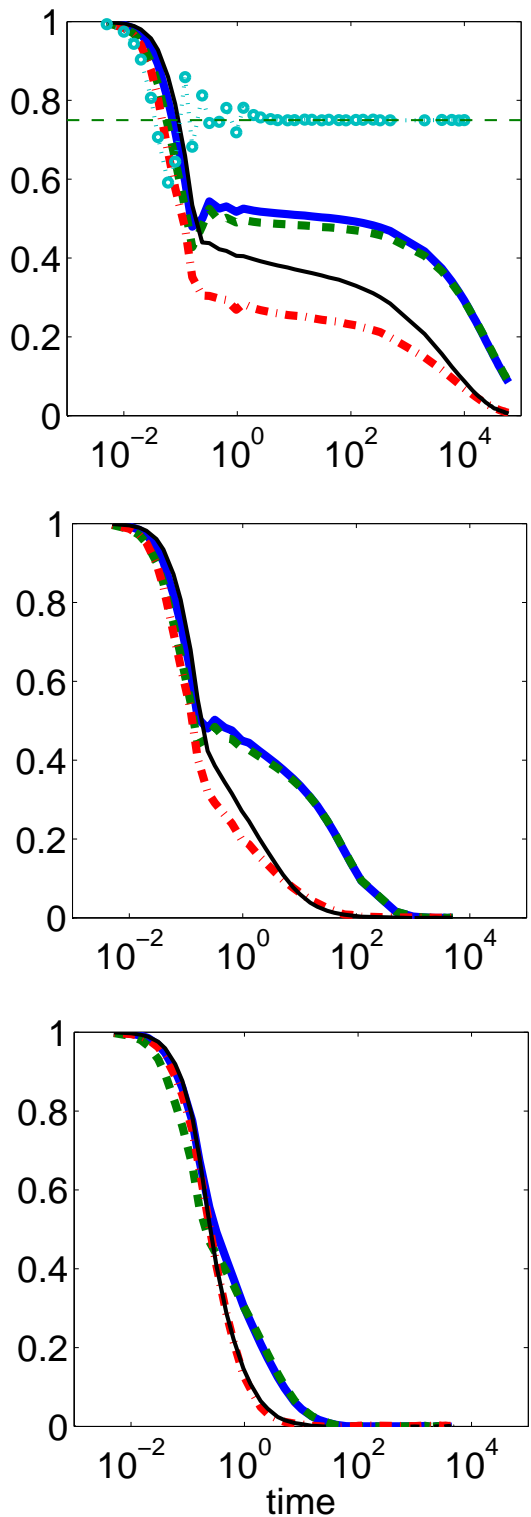}}
  \caption{(color online) For $T=0.38$ (top), $T=0.50$ (middle) and $T=1.00$ (bottom), is shown the autocorrelation of various quantities related to the polarisation. Solid blue: $\lan \bm\tau_j(0)\cdot\bm\tau_j(t)\ran/\lan \bm\tau_j^2\ran$; Dashed green: $ \lan \hat{\bm\tau_j}(0)\cdot\hat{\bm\tau_j}(t)\ran$ with $\hat{\bm\tau}_j=\bm\tau_j/|\bm\tau_j|$; Dash-dotted red: $\lan \de|\bm\tau_j|(0)\de|\bm\tau_j|(t)\ran/\lan(\de|\bm\tau_j|)^2\ran $ with $\de|\bm\tau_j|=|\bm\tau_j|-\lan|\bm\tau_j|\ran$. The thin black curve is the normalized autocorrelation of $U_j$ from Fig. \ref{autocorr_UU}. In the top figure, the blue circles correspond to rescaled and shifted autocorrelation of the total force acting on a monomer, the zero level being the thin dashed  line. This curve shows that the relaxation of the force is much faster than that of the polarisation.}
\label{autocorrtau}
\end{figure}

As for the relaxational dynamics of $\bm\tau_j$,  Fig. \ref{autocorrtau} reveals a behaviour similar to what one observed for $U_j$.
After a rapid initial decay, a plateau regime is reached, followed by the final $\al$ relaxation. What is worth stressing is that the plateau corresponding to the relaxation of the modulus of $\bm\tau_j$ is smaller than that of the total and directional relaxation. The relaxation of the modulus of $\bm\tau_j$ proceeds  faster though at a pace comparable to that of the volume relaxation. On the other hand, the total or directional relaxations are much more blocked in the plateau regime. One can probably attribute this behaviour to the fact already touched upon, namely that the polarisation is sensitive to the connectivity in our model: As a result, the slower behaviour of the directional relaxation for $T=1$ may be attributed to polymeric effects, as reflecting the presence of a slower relaxation time (Rouse time) necessary for the polymer to relax its orientation (actually only a blob relaxation time would be relevant here, but for $N=4$ the blob and Rouse times are one and single concept). Therefore, the blatant shoulder for $T=0.50$ and high plateau for $T=0.38$ for the directional relaxation (or total relaxation: it is quite clear from the curves that the  relaxation of the total polarisation is impeded mainly by that of its direction) accounts for a reorientational polymeric relaxation time, already slower in normal conditions, and considerably increased by the slowing down of the monomeric dynamics \footnote{Notice that a tailoring of the system so as to adjust the mean distance between intrachain neighbours with that of Voronoi nearest neighbours is just impossible, because such a system would partially crystallise on the monomeric scale (not with respect to the chain backbone).} . A systematic study of this issue with respect to $N$ would be clearly desirable.

\medskip

As we have already noticed, polarisation and force are structurally correlated. However, dynamically speaking, their behaviour is very different, insofar as the force cannot be correlated in the long run (this  would imply indeed drifts in the motion of particles), see Fig. \ref{autocorrtau} (top). As for the volume $U_j$ (see the discussion at the end of the subsection \ref{DBP}), the interruption of the relaxation of $\bm\tau_j$ (at the early beginning of the correlation plateau) should account for the structural disorder of the inherent structures around which the system fluctuates in this early post-microscopic time regime. Notice that as far as the  IS is concerned, the force field is correspondingly irrelevant, because it is everywhere zero.

\subsection{Correlation polarisation/displacement}

\begin{figure}
\centering\resizebox{7cm}{!}{\includegraphics{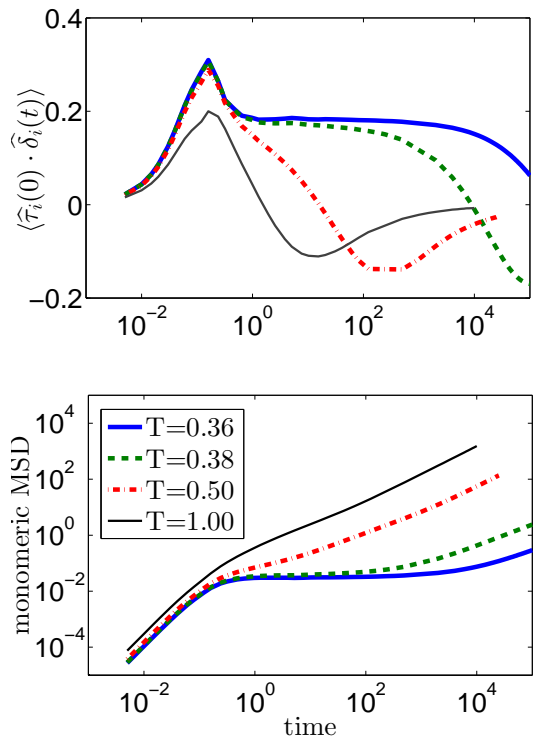}}
  \caption{(color online) Top: Correlator $\lan \hat{\bm \tau_i}(0)\cdot\hat{\bm \de_i}(t)\ran$ as a function of time for several temperatures ($\hat{\bm X}=\bm X/|\bm X|$). Bottom: Monomeric mean squared displacement for the same temperatures.}
\label{corrpoldis}
\end{figure}

As the polarisation vector $\bm\tau_i$  of the particle $i$ is structurally correlated to the instantaneous total force $\bm F_i$ acting on $i$, necessarily a correlation builds up  between the initial value of $\bm\tau_i$ and the displacement $\bm\de_i(t)=\bm r_i(t)-\bm r_i(0)$ of $i$ in the short time regime. This can be seen in fig. \ref{corrpoldis} (top), where the correlator $\lan \hat{\bm \tau_i}(0)\cdot\hat{\bm \de_i}(t)\ran$ is plotted versus time (we note $\widehat{\bm X}=\bm X/|\bm X|$).
 Clearly a noticeable correlation develops during the ballistic regime (see the mean-square displacement (MSD) in the bottom of Fig. \ref{corrpoldis}), crosses over a maximum, and then decreases.  For intermediate times the decrease is temporarily blocked at a plateau, which is the more pronounced, the lower the temperature. It is again possible to interpret this interrupted decorrelation within the scenario of an intermediate timescale dynamics based on the hopping between adjacent IS \cite{Schroderetal}. In the early timescales of the plateau relaxation, only a few proportion of particles have actually moved out of their local cage (see Fig. \ref{fractionofmobile}). As a result, a large part of the signal should come from particles which are stuck, rattling back and forth in the cage of their neighbours. The plateau of Fig. \ref{corrpoldis} (top) is thus mainly the imprint of particles having relaxed to a long-lived effective cage associated to an IS.

\begin{figure}
\centering\resizebox{7cm}{!}{\includegraphics{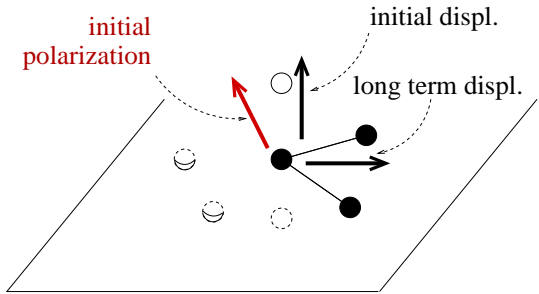}}
  \caption{Why the correlator of the fig. \ref{corrpoldis} changes sign in the long time regime, see text for details.}
\label{sketchpol}
\end{figure}

\begin{figure}
\centering\resizebox{7cm}{!}{\includegraphics{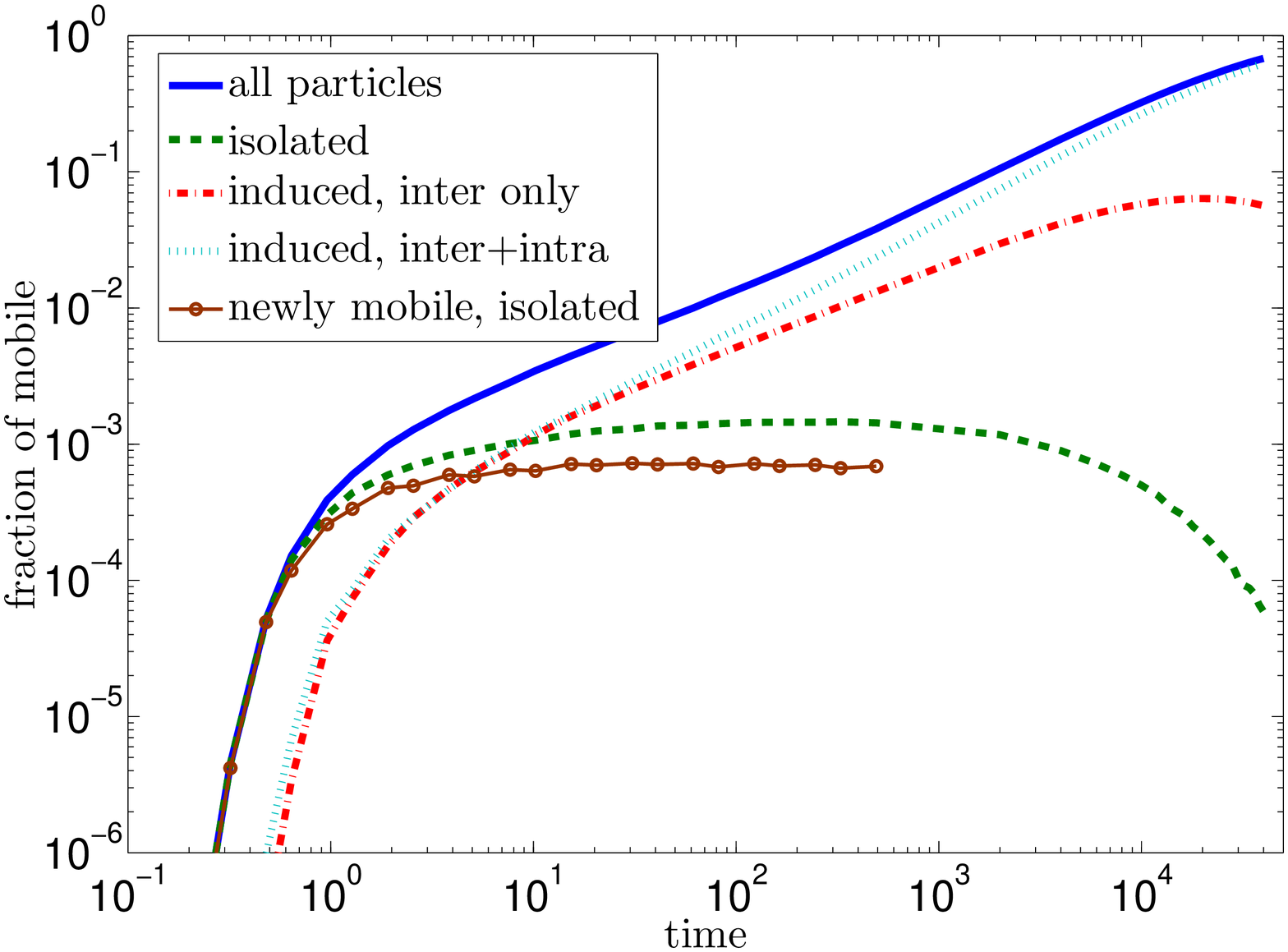}}
  \caption{(color online) Fraction of mobile particles for $T=0.38$ and $r_c=3\lan \de_i^2\ran=0.5765$ (solid blue). Note that the concept of ``mobile'' and ``immobile'' depends on the time considered. The other curves are the fraction of mobiles particles fulfilling an additional criterion: (i) dashed green (``isolated''): mobile particles whose initial neighbours are {\em all} immobile at time $t$. (ii) dash-dotted red (``induced, inter only''): mobile monomers $i$ with one or several mobile neighbours $j$, neither of the monomers $j$ belonging to the same molecule as that of $i$. (iii) dotted pale blue (``induced, inter+intra''): mobile monomers with mobile neighbours, not belonging to the preceding category. (iv) brown circles and solid line (``newly mobile, isolated''): particles of the same breed as (i), with the additional criterion that  ``newly mobile'' particles at time $t$ must have become mobile between the last time step and $t$ only ---the circles are placed on the time steps at stake. This last curve stops early for technical reasons only.}
\label{fractionofmobile}
\end{figure}

A peculiar feature of the late relaxation of the polarisation/displacement correlation is the marked and long anticorrelation which occurs for a typical time similar to the $\al$ relaxation time. This again is a peculiarity of our model glass-forming liquid of short oligomers: To understand why this is so, consider an initial situation like that depicted in fig. \ref{sketchpol}. 
In this figure, the central black monomer is represented with its two intrachain neighbours, and the plane that these three particles define is outlined. If one assumes a locally icosahedral order, a pentagon of neighbouring particles can be drawn in that plane (for sake of clarity, of the three remaining  in-plane neighbours needed to complete the pentagon, only two have  been drawn ---with cupped bottoms to highlight their location across the plane). This in-plane pentagon is in average slightly distorted due to the intra-chain bond lengths which are a bit shorter that the average nearest-neighbour distance. In the figure, the two remaining monomers represent roughly the barycenters of the neighbours located above (solid) and below (dashed) the plane. The configuration has been chosen so that the particles below are typically closer to the central particle than that located above. This fluctuation plus the systematic distortion of the in-plane pentagon make the  polarisation associated to that configuration (denoted ``initial'' in the figure) going from bottom to top and tilted to the left. The short time displacement will correspond to the smoothing of the vertical fluctuation, which is associated with a force fluctuation, whereas no in-plane relaxation takes place preferentially from right to left, because the distortion of the in-plane is structural. Therefore the short term displacement is mainly vertical and its scalar product with the initial polarisation is positive. For longer times, the displacement of the monomer at stake (but for a time scale still smaller than the diffusion time) corresponds to the polymeric relaxation of the different internal degrees of freedom. As a result, the late motion of the monomer  corresponds to the exploration of the space defined by the blob of the three neighbours drawed in black, and therefore will be preferentially towards the right, leading to a negative scalar product. 

\medskip

The correlator of fig. \ref{corrpoldis} is somewhat too general to describe acurately the relation between polarisation and displacement when the temperature is so low that the dynamics becomes heterogeneous. To get a glimpse on how the displacement decorrelates from polarisation differentially between mobile and stuck regions, one first defines what a mobile particle is: As a rule, a monomer $i$ is termed `mobile' at time $t$ if one has $|\bm\de_i(t)|>r_c=3 \sqrt{\lan \bm r_{\rm pl}^2\ran}$ where $\lan \bm r_{\rm pl}^2\ran$ is the plateau value of the monomeric mean-square displacement (this value for $r_c$ corresponds to the usual criterion used to qualify the subpopulation of mobile particles at a time corresponding to the maximum of the nonergodicity parameter \cite{KobDonatiPRL1997,BaschnagelVarnik}). Among the population of mobile particles, it is clear that those whose displacement has been provoked by the motion of a neighbouring particle are not expected to sustain a correlation between polarisation and displacement directions. On the contrary, these induced motions are likely to be directed toward the vacancy left by the initially moving particle (string-like motion). Therefore, one is naturally led to distinguish two subpopulations among the mobile particles at time $t$: 1) A minority fraction, termed thereafter the dynamical seeds, which has substantially moved at time $t$, whereas their immediate environment at time $t=0$ (the neighbouring particles sharing a Voronoi facet with the particle under consideration) has not, and 2) the remaining ones, whose motion can be considered roughly as  induced by the prior motion of a neighbouring monomer. Of course, this distinction is sensible only for sufficiently supercooled systems, where the motion of particles is substantially heterogeneous.
This distinction is illustrated in figure \ref{fractionofmobile}, where the mean fraction of mobile particles is plotted in solid blue. Obviously, this curve is increasing and saturates at 1 for very long times (not shown). The intermediate behaviour is approximately a power law with an exponent around $2/3$ (this is largely coincidental, for it is not robust with respect to a variation of the choice of $r_c$). Together with this curve is shown also the fraction of dynamical seeds (dashed green curve). After the short time regime, where this curve is obviously superposed with the total curve, the fraction of dynamical seeds becomes roughly constant for a very long time. This should come from a balance between two fluxes, one providing isolated mobile particles, the other removing particles from this category: On the one hand, some immobile particles within an immobile environment become progressively mobile, probably because they are not  too far from a reorganizing region. On the other hand, some dynamical seeds having initiated neighbouring displacements consequently leave the category of isolated mobile particles. For sake of completeness, Fig. \ref{fractionofmobile} displays also the tiny flux of ``newly mobile'' isolated particles (line+circles), i.e. particles mobile at time $t$ but immobile at all prior recording time steps (these recording time steps are shown by the symbols)

The very large discrepancy between the total fraction of mobiles particles and the fraction of isolated mobile particles, as soon as the early microscopic regime is overcome, witnesses the important fact that most of the subsequent diffusion events occurs by cascade, a dynamical seed inducing a neighbouring motion, which in its turn provokes another one, and so on. As our system is made of oligomers, we tested in Fig. \ref{fractionofmobile} also the effect of connectivity on the cascade. The red dash-dotted curve is the fraction of  mobile particles having at least one mobile neighbour, but this or these mobile neighbour(s) do(es) not belong to the same molecule. The light blue dotted curve is the complementary fraction of not isolated mobile particles, i.e. mobile particles with at least a mobile neighbour belonging to the same molecule. These two curves are initially very close, which indicates that the connectivity enhances from the beginning the probability that a mobile monomer induces the motion of a neighbouring monomer belonging to the same chain. This point comes from the fact that a given monomer has typically 12-14 neighbours, only 3 of them at best belong  to the same molecule (since our system is made of oligomers of length 4). For late times, the red dash-dotted curve becomes obviously negligible, since the vast majority of mobile particles  (i) are not isolated, and (ii) have an environment with multiple, intra and inter-chain mobile monomers.

\medskip

\begin{figure}
  \centering \resizebox{7cm}{!}{\includegraphics{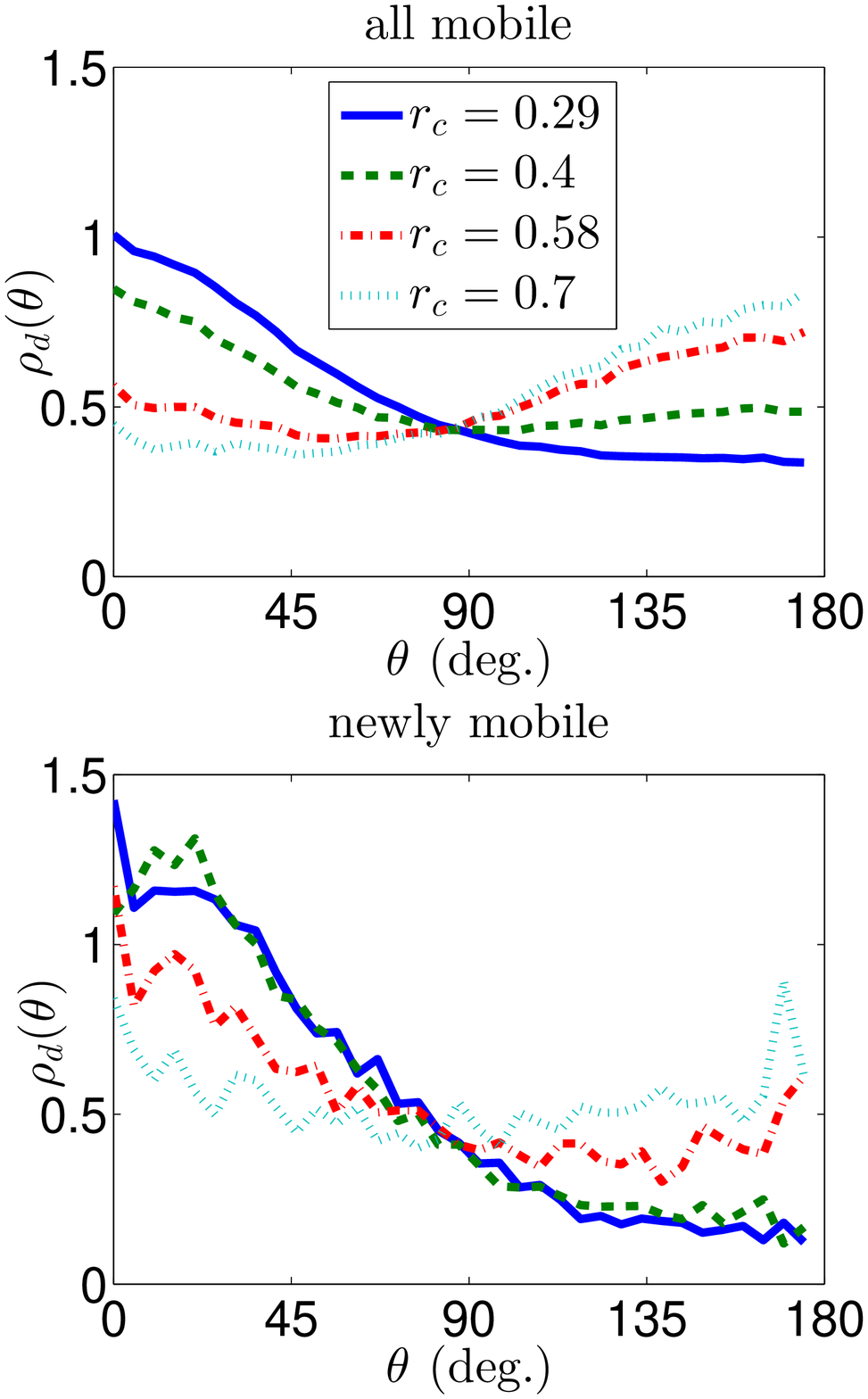}}
  \caption{}
\label{vivid}
\end{figure}

If one comes back to the correlation of polarisation and displacement, it is clear that this correlation disappears at once for all diffusion events triggered by a displacement of a neighbouring particle. As a result, most of the plateau in Fig. \ref{corrpoldis} is built up by stuck particles which are unable to move beyond the initial release of the force fluctuation into the local basin of the inherent structure associated to the initial condition. A vivid illustration of this is obtained if one analyses the polarisation/displacement correlations of the small fraction of ``newly mobile'' particles at a time which is already considerable. In fig. \ref{vivid} (top) one sees the angular density $\rho_d(\theta)$
(i.e. the probability density divided by $\sin\theta$), of the angle between $\bm\tau_i(0)$ and $\bm\de_i(t)$, for a large time $t\sim 123$ and various $r_c$. Clearly, if $r_c$ is too high, the `mobile' particles are indeed mobile thanks to the accumulation of several jumps between IS. It is therefore natural to observe a relatively flat angular distribution (the probability excess for angles near $180^\circ$ is again an outcome of the connectivity). For smaller values of $r_c$, the curves are mainly  coding the already mentioned freezing of the initial relaxational moves due to the temporary caging around the inherent structures. It is worth stressing that we have no explanation for the apparent common value of the different curves for $\theta\sim 90^\circ$.

What is new is the bottom figure, where the same distributions are plotted, but calculated among the tiny subpopulation of ``newly mobile'' particles. Again, if $r_c$ is too large, the ``recent'' substantial moves are actually the result of several IS rearrangements and the correlation with the initial polarisation direction is lost. But, for $r_c\lesssim 0.4$, one observes that the ``fresh'' move of the particle (which occurred here at a time  $t\in[82,123]$) has kept a strong correlation with the initial polarisation, despite a quite long waiting time before the move. We checked that the curves associated to the newly mobile particles are for $r_c\lesssim 0.4$ almost independent of time in the whole plateau time region. The result is quite interesting, for it shows that (i)  still regions are able to preserve for quite long times the memory of the initial polarisations, and (ii) the substantial motion of a particle which can be considered as arising from a single dynamical hopping (and not a series of correlated events) is quite small (here $<0.4$), definitely smaller than the value usually taken to define the subpopulation of mobile particles \cite{KobDonatiPRL1997} (which is $\sim 3 \lan \bm r_{\rm pl}^2\ran\sim 0.58$ in our system).

\subsection{Polarisation torque}

The system of vectors $\bm\tau_j$, like the force field, sums up to zero (with the important difference that one cannot decompose each $\bm\tau_j$ into vectors which would obey a kind of third Newton law). Consequently, the polarisation torque defined by
\begin{figure}
  \centering\resizebox{6cm}{!}{\includegraphics{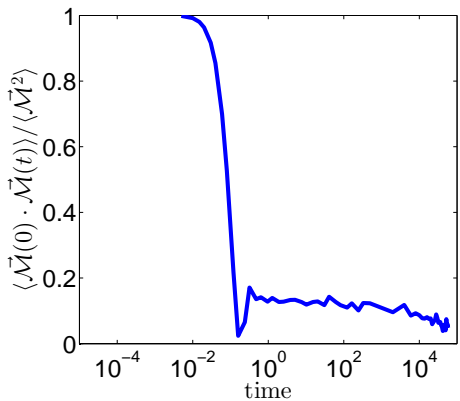}}
  \caption{Normalized autocorrelation of the polarisation torque for $T=0.36$.}
  \label{autocorr_torque}
\end{figure}
\begin{align}
  \bm{\mcal{M}}&=\sum_j \bm r_j\times \bm\tau_j
\end{align}
does not depend on the choice of the origin. For each configuration, this torque accounts for a geometrical global anisotropy. In liquid conditions, this anisotropy is reshuffled within a microscopic time scale, but for supercooled states, it   retains some correlation until the $\al$ relaxation. This is shown in Fig. \ref{autocorr_torque} for the temperature $T=0.36$ (the other temperatures are not shown due to a lack of statistics on this global variable). Again, the plateau is associated  to the fact that a typical inherent structure has a nonzero value of its torque, that the hopping from IS minimum to IS minimum only progressively decorrelates the torque from its initial orientation and magnitude. It is likely that this large-scale anisotropy couples to the shear properties of the system, and makes any rheological response on time scales faster than the $\al$ relaxation time dependent on the direction of strain.

\subsection{Polarisation field}
\medskip

\begin{figure}
\centering  \resizebox{7cm}{!}{\includegraphics{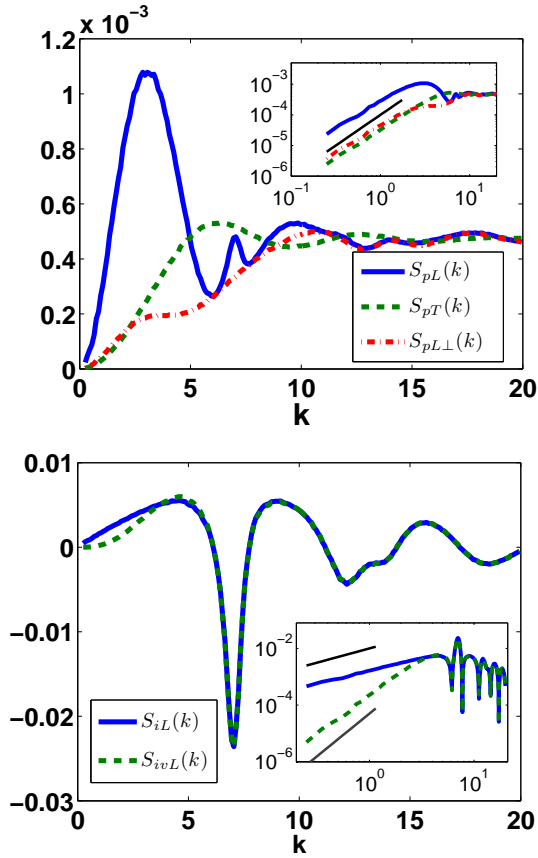}}
  \caption{(Color online) Top: Longitudinal $S_{pL}(k)$ (solid blue) and transverse $S_{pT}(k)$ (dashed green) fluctuations of the geometrical polarisation $\bm \tau_j$; The function $S_{pL\perp}(k)=N^{-1}\lan p_{L\perp}(-\bm k)p_{L\perp}(\bm k)\ran$ is also shown (red dash-dotted). The inset shows the equivalent log-log plot to highlight the powerlaw $k^2$ (black thin line). Bottom: Cross-correlations $S_{iL}(k)$ (solid blue) and $S_{ivL}(k)$ (dashed green). The inset shows the log-log plot of their moduli, together with the powerlaw $k^1$ and $k^3$ (thin top and bottom lines respectively).}
  \label{SpLSpT}
\end{figure}

\begin{figure}
\centering  \resizebox{7cm}{!}{\includegraphics{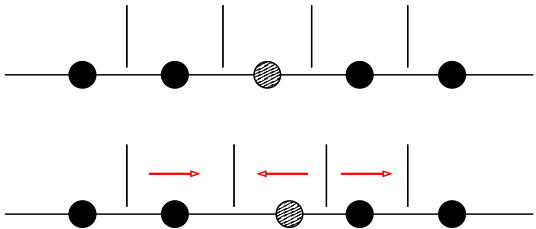}}
  \caption{(Color online) Origin of the first peak in $S_{pL}(k)$. On top is sketched the average position of the particles around a marked (shaded) one; The vertical lines show the boundaries of the corresponding Voronoi cells. At the bottom is shown a position fluctuation of the marked particle; it induces nonzero polarisation vectors (red arrows of arbitrary lengths) in the immediate vicinity of the particle, with a positive correlation between the vectors surrounding the marked particle vector.}
  \label{petitschema}
\end{figure}

\begin{figure}
\centering  \resizebox{7cm}{!}{\includegraphics{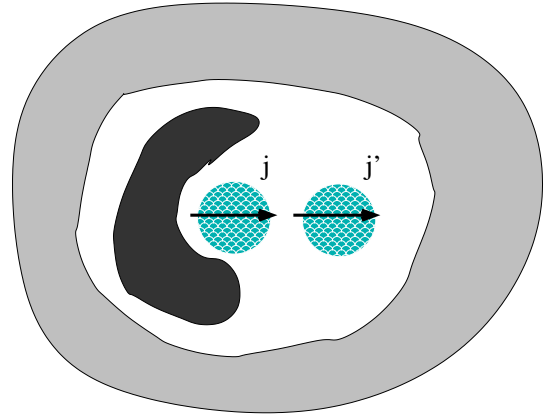}}
  \caption{Positive correlation of polarisation between two nearest neighbour particles (blue circles). The positive density fluctuation left hand the particle $j$ (dark zone) induces a polarisation $\bm\tau_j$ toward the right; this density fluctuation is surrounded by a negative density correction (light zone), which encompasses $j'$ and yields a polarisation $\bm\tau_{j'}$ pointing the same direction as $\bm\tau_j$.}
  \label{whycorrelpositive}
\end{figure}

As for the volume fields, we  also study  the autocorrelation functions of $\bm p$ with itself. As is usual for e.g. the current distribution function of the liquid physics, we proceed by defining in the Fourier space the components of $\bm p(\bm k)$ transverse and longitudinal with respect to the direction $\hat{\bm k}=\bm k/k$:
\begin{align}
  p_L(\bm k,t)&\equiv -i(\bm p(\bm k,t)\cdot \hat{\bm k})\\
\bm p_T(\bm k,t)&\equiv \bm p(\bm k,t)-i p_L(\bm k,t)\hat{\bm k}\\
\text{with  }\ \bm p(\bm k,t)&\equiv\sum_j\bm\tau_j(t)\exp(i\bm k\cdot\bm r_j(t))
\end{align}
(The $\sqrt{-1}$ in the definition of $p_L$ is just for $k$-inversion symmetry convenience). The (static) autocorrelation function of these longitudinal and transverse components are  defined by 
\begin{align}
  S_{pL}(k)&=N^{-1}\lan p_L(-\bm k) p_L(\bm k)\ran\\
  S_{pT}(k)&=(2N)^{-1}\lan\bm p_T(-\bm k)\cdot\bm p_T(\bm k)\ran
\end{align}
and are plotted in Fig. \ref{SpLSpT} (top). 
The scale of the vertical axis, homogeneous to a squared length, shows that the typical values of $\tau_j$ are only $3-4\%$ of the nearest-neighbour distances, which is again a signature of the typically small fluctuations of the local structural properties in the  dense fluid regime. The longitudinal part is rather structured with a first peak for $k<5$, whose qualitative origin is explained in Fig. \ref{petitschema}
for a one-dimensional system. The secondary peak at $k=k_c$ is however not accounted for by the simple 1D representation. On the contrary, for a 1D model with excluded volume, the first structural peak corresponds to a minimum of $S_{pL}(k)$ (not shown), which is intuitively understood from the Fig. \ref{petitschema}, where a maximum anticorrelation is expected between the $\tau_j$ of nearest neighbours. As a result, this secondary peak can be  related to some structural correlation involving the transverse directions (see Fig. \ref{whycorrelpositive}): Let us consider that $k\sim k_c$ is fixed. This corresponds to selecting a characteristing length $\sim k_c^{-1}$ as a filter. The particle pairs, say $(j,j')$ which contribute to $S_{pL}(k)$ are those at a distance $\sim k_c^{-1}$, i.e. the nearest neighbour pairs. From pair to pair, the two polarisation vectors $\bm\tau_j$ and $\bm\tau_{j'}$ are fluctuating because of the varying environment. By the way a large part of the immediate vicinities of $j$ and $j'$ are only weakly correlated, which induces a relatively weak correlation of $\bm\tau_j$ and $\bm\tau_{j'}$. Nevertheless, a positive correlation along the particle axis is expected, due to a density dipole effect: if (cf. fig. \ref{whycorrelpositive}) on the ``external'' side of $j$ a positive fluctuation of density is observed, the polarisation $\bm\tau_j$ is typically oriented toward $j'$ since the distance between $j$ and $j'$ is precisely assumed to have the average distance $k_c^{-1}$. But in a fluid, the correlations are very short ranged, which compels the positive density correlation to be immediately surrounded by a negative one (to restore the mean density beyond a distance $\sim 2-3\ k_c^{-1}$). This density depletion does not affect the distance $j-j'$ by assumption, but will typically provide the outward vicinity of $j'$ with a density slightly lowered, whence a $\bm\tau_{j'}$ pointing the {\em same} direction as $\bm\tau_j$.
This discussion shows clearly that $p_L$ is sensitive to the local details of the density field, and notably to the transmission of the correlation due to excluded volume effects. Moreover, one shows that the reasoning involves clearly more than two particles, which witnesses the typical many-body nature of the Voronoi-inspired observables.

\medskip

As regards the transverse correlator $S_{pT}$, it shares with $S_{pL}$ a departure $\propto k^2$ for low $k$ (see inset of Fig. \ref{SpLSpT} top), which can be shown along lines similar to those developped in appendix \ref{App_pol_1}. In the microscopic domain $k\simeq k_c$, the function reaches a plateau, only weakly structured; The plateau value is $\lan\bm\tau_j^2\ran/3$, the same as $S_{pL}$.

\medskip

The longitudinal polarisation field $p_{L}(\bm k)$ deserves some more comment. First, the relation \myref{Svkapp} does not imply that $S_{pL}(k)$ and $S_{v}(k)/k^2$ are equal in the limit $k\rightarrow0$. These two quantities are indeed proportional to $k^2$, but a priori with  different prefactors. However, Eq. \myref{Svkapp} explicitely indicates that $p_{L}(-\bm k)$ can couple to $\rho(\bm k)$ and $\rho_v(\bm k)$, i.e. that
\begin{align}
  S_{iL}(k)&\equiv N^{-1}\lan p_L(-\bm k)\rho(\bm k)\ran\\
  S_{ivL}(k)&\equiv N^{-1}\lan p_L(-\bm k)\rho_v(\bm k)\ran
\end{align}
are both non zero (this is not the case with the transverse field, which does not couple to $\rho$ or $\rho_v$, due to incompatible symmetries). They are respectively $\propto k^1$ and $\propto k^3$ for vanishing $k$ and go to zero for high $k$ due to the isotropy, as can be seen in Fig. \ref{SpLSpT} (Bottom). As for $\rho_v$ and $\rho_{v\perp}$, we define  an orthonormalized coordinate $p_{L\perp}$ from $p_L$, and its correlator $S_{pL\perp}(k)$ by
\begin{align}
  p_{L\perp}(\bm k)&=p_L(\bm k)-\frac{S_{iL}(k)}{S(k)}\rho(\bm k)\\
S_{pL\perp}(k)&=N^{-1}\lan p_{L\perp}(-\bm k)p_{L\perp}(\bm k)\ran\nonumber\\
&=S_{pL}(k)-\frac{S_{iL}(k)^2}{S(k)}
\end{align}
The structure factor $S_{pL\perp}(k)$ is plotted in Fig. \ref{SpLSpT} (top). The orthonormalization procedure has the drastic effect of removing the main parts of the structuration peaks. Interestingly, the shoulder at $k\sim 3$ is a remnant of the large former peak, and accounts for the specific three- or higher-point correlation ``content'' of the physics described schematically in fig. \ref{petitschema}. This new  characteristic length provided by $S_{pL\perp}$ is specific to the polarisation, and such a shouldering is absent in the corresponding structure factor $S_{v\perp}$ associated to the orthogonalized volume field.

\section{Conclusion}

In this paper we put forward two new  fields for the description of particle assemblies,  defined thanks to the Voronoi tessellation of the configurations. The first one, the volume field $\rho_v(\bm r)$, associates to each Dirac function at the particles location a weight proportional to the Voronoi cell volume. 

We have shown that the large scale fluctuations of this field display an anomaly that we proved to be related to a very peculiar feature of the Voronoi partitions, not noticed up to now to the best of our knowledge: The geometrical polarisation $\bm\tau_j$ of the cell $j$, a vector accounting roughly for the local anisotropy of the microscopic arrangement, is a conserved vectorial field, i.e. all geometrical polarisations sum up to zero. The associated polarisation field $\bm p(\bm k)$ is in our opinion a promising tool, because it has the unique property of being a conserved field bearing a vectorial information about the structure of the system. 

We gave thereafter the main properties of the individual geometrical polarisations (we did not dwell too much on the individual Voronoi volumes, because their properties for dense fluids are rather well known \cite{Glotzer_Voronoi}). These vectors are approximately described by  a Maxwell-Boltzmann statistics, and are statically somewhat correlated to the force field. In spite of this correlation, the dynamical behaviour of $\bm F$ and $\bm \tau$ are drastically different for supercooled liquids, and we think that these fundamental discrepancies can be traced back to the typical microscopic disorder of the inherent structures around which the supercooled liquids are temporarily oscillating during early microscopic regime (from $t=0$ until the very beginning of the plateau regime). These IS have a nonzero polarisation field, preventing the actual polarisation to decorrelate fully before the $\al$ process (these arguments apply to the volume field as well). A natural issue we will address in a future publication is to what extent the  diffusion of the system in the phase space, on time scales large enough as to allow an  interpretation in terms of hopping or shifting from IS to IS, is statistically correlated to the residual polarisation and volume fields of the IS.  
The processes by which these relaxations take place have been shown to be very peculiar \cite{Schroderetal}, involving for instance highly anisotropic clusters of mobile particles. It may occur that these clusters are somehow related to the clusters of particles one can readily define from the IS residual polarisation field by lumping together particles with the rule that two neighbouring particles belong to the same cluster if the polarisation of one is pointing mostly to the other or vice-versa. Designed from a configuration polarisation field, one may think that these basic clusters are somehow a blurred version of a soft mode, owing to the structural polarisation/force correlation. Designed instead from a residual polarisation field of an IS, they could perhaps be correlated to the reaction path in phase space, i.e. to the path in phase space followed by the system so as to minimize the barrier from one IS to the next.

Finally, it is worth noting that $\rho_v(\bm k)$ and the longitudinal part of the polarisation field $p_L(\bm k)=-i \bm p(\bm k)\cdot \bm k$ share the same symmetries as the density field itself. As a result, they are potential candidates to devise an extended mode-coupling theory, enlarged so as to consider as relevant variables not only $\rho(\bm k)$ but also the correctly orthogonalized parts of $\rho_v$ and $p_L$. This extension would provide the ideal MCT with degrees of freedom aiming at describing the structure on a more real-space, integrated basis. The actual realization of such an extension relies on the possibility to express frequencies, vertices, etc\ldots involving variation of Voronoi volume and polarisation in terms of geometric quantities amenable to simple numerical calculations. The simplicity of the Voronoi construction allows such explicit simple expressions, an example of it being eq. \myref{gradU}. Actually, the major challenge for such a program resides mostly in the evaluation of vertices, which must be precise and simple. For the MCT, these requirements are met thanks to the factorization Ansatz by which the vertices are simple formulas of the structure factor only. The price to pay toward an enrichment of the ideal theory is a relative loss of simplicity. A certain amount of work has still to be done to make the complexification as manageable as possible.

\section{Appendices}

\subsection{Appendix : The gradient formula for a Voronoi cell}\label{AppMakse}

In this appendix, we show that when two distinct particles $i$ and $j$ in a configuration are ``nearest-neighbours'', that is their Voronoi cells share a common facet $S_{ij}$ ($S_{ij}$ names the facet or terms its area, according to the circonstance), one has
\begin{align}
  \nab_j U_i&=-\frac{S_{ij}}{r_{ij}}\bm s_{ji}\label{nabjUi}
\end{align}
where $r_{ij}=|\bm r_j-\bm r_i|$, and $\bm s_{ji}$ is the vector from the particle $j$ to the barycenter of the facet $S_{ij}$. To prove this formula, we use the following formula \cite{Makse}:
\begin{align}
  U_i&=\frac{1}{d}\oint d\Om [L_i(\bm n)]^d\\
  L_i(\bm n)&=\min_{j/\bm n\cdot\hat{\bm r_{ij}}>0}\frac{r_{ij}}{2\bm n\cdot\hat{\bm r}_{ij}}\label{Lidef}
\end{align}
In this formula, valid for a Voronoi cell in any dimension $d\geq 2$, the integration is performed over the angular directions viewed from the position of the particle $i$, $\hat{\bm r}_{ij}$ stands for $\bm r_{ij}/r_{ij}$, and $\bm n$ is a unit vector pointing in the direction $\Om$, and $L_i(\bm n)$ is precisely the distance between $i$ and its Voronoi cell boundary in the direction $\bm n$.

\medskip

We are interested in computing derivatives of $U_i$ with respect of coordinates. To this end, it is useful to note that
\begin{multline}
  \min_{\text{positive items}} \{A,B,C,(\ldots)\}=\\
A \Te(A) H(B,A)H(C,A)
+B\Te(B)H(A,B)H(C,B)\\+C\Te(C)H(A,C)H(B,C)
\end{multline}
with
$H(B,A)\equiv1-\Te(B)\Te(A-B)$, 
where $\Te$ is the Heaviside function. Now, if one derives with respect to a variable $a$ to which $B$ and $C$ are independent, we get, due to the assumed continuity of the functions
\begin{align}
  \pa_a\min_{\text{positive items}} \{A,B,C,(\ldots)\}&=(\pa_a A)\Te(A) H(B,A)H(C,A)\label{paa}
\end{align}
There is however a proviso with this last formula : we assumed that the variables $A,B,$etc\ldots are never zero. When a degeneracy between the particles is possible (exact superposition of two of them), some singularities are likely to show up. As it never happens with particles with  excluded volume, we disregard these mathematical limiting cases. Formula \myref{paa} allows a controlled derivation of $U_j$: we have rigorously
\begin{align}
  \bm\nab_jU_i&=\frac{1}{2^d}\int_{S_{ij}}d\Om\frac{r_{ij}^{2(d-1)}}{(\bm n\cdot\bm r_{ij})^{d+1}}[2\bm r_{ij}(\bm n\cdot\bm r_{ij})-r_{ij}^2\bm n]
\end{align}
Now we consider only the three dimensional case $d=3$. We have here $d\Om=(d\bm S\cdot \bm n)/r^2=4dS(\hat{\bm r}_{ij}\cdot\bm n)^3/r_{ij}^2$ whence
\begin{align}
  \bm \nab_j U_i 
&=
\demi \int_{S_{ij}}dS\frac{\hat{\bm r}_{ij}(\bm n\cdot \hat{\bm r}_{ij})-\bm n^\parallel}{\hat{\bm r}_{ij}\cdot \bm n}
\end{align}
where $\bm n^\parallel=\bm n-(\bm n\cdot\hat{\bm r}_{ij})\hat{\bm r}_{ij}$ is the component of $\bm n$ parallel to $S_{ij}$. The final step is to ``change the viewpoint'', i.e. replace the unit vector $\bm n$, adapted to a line of sight issued from the particle $i$ toward the surface $S_{ij}$,  with the unit vector $\bm n'$ corresponding to a sight issued from $j$. The vector $\bm n'$ is the symmetrical of $\bm n$ with respect to the plane containing $S_{ij}$. We get
\begin{align}
  \bm\nab_j U_i&=-\demi \int_{S_{ij}}dS\frac{\hat{\bm r}_{ji}(\bm n'\cdot \hat{\bm r}_{ji})+(\bm n') ^\parallel}{\hat{\bm r}_{ji}\cdot \bm n'}\\
&=-\demi \int_{S_{ij}}dS\frac{2r'\bm n'}{r_{ij}}=-\frac{S_{ij}}{r_{ij}}\bm s_{ji}\label{eq21}
\end{align}
(where $r'\bm n'$ is the running vector from $j$ to a point of $S_{ij}$). It is important to stress that in general $\bm s_{ji}\neq -\bm s_{ij}$, whereas one has obviously $\bm r_{ij}\equiv \bm r_j-\bm r_i=-\bm r_{ji}$. Notice that  \myref{eq21} has the simple corollary
\begin{align}
  \bm \nab_j U_i-\bm \nab_i U_j&=S_{ij}\hat{\bm r}_{ij}
\end{align}

\subsection{Appendix : The $k^4$ behaviour of $S_v(k)$}\label{App_pol_1}

(We deal here with 3D systems, but the generalization to $n\geq 2$ dimensions is obvious). We consider first a slight variant of $S_v(k)$, by defining $\tilde{S}_v(k)=N^{-1}\lan \tilde{\rho}_v(-\bm k)\tilde{\rho}_v(\bm k)\ran$ with
\begin{align}
  \tilde{\rho}_v(\bm k)&=\rho_v(\bm k)-K(k)v^{-1}\int d^3r e^{i\bm k\cdot \bm r}\\
&=\sum_j e^{i\bm k\cdot\bm r_j}\underbrace{v^{-1}\left(U_j-K(k)\int_{U_j}d^3r e^{i\bm k\cdot [\bm r-\bm r_j]}\right)}_{\equiv A_j(k)}
\end{align}
where $K(k)$ is a function, with the sole requirement that $K(0)=1$, chosen in such a way that $\lan A_j(k)\ran=0$ for all $k$, that is
\begin{align}
  K(k)&=\frac{v}{\left\lan\displaystyle\int_{U_j}d^3r e^{i\bm k\cdot [\bm r-\bm r_j]}\right\ran}
\end{align}
Notice that $K(0)=1$ as it must, and that the denominator is a kind of generating function for the inertial moments of the Voronoi cell. 
It is readily seen that $S_v(k)$ and $\tilde{S}_v(k)$ are identical, but for $k=0$. 

$A_j(k)$ has a {\em bona fide} Taylor expansion  which is, up to the first order
\begin{align}
  A_j(k)=-i\bm k\cdot\bm \tau_j+o(\bm k)\label{eqAA1}
\end{align}
We have besides
\begin{align}
  S_v(k)&=N^{-1}\sum_{j,j'}\lan A_j(k)A_{j'}(-k)\exp(i\bm k\cdot[\bm r_j-\bm r_{j'}])\ran\label{Svkapp}
\end{align}
and, on the assumption that $A_j$ and $A_{j'}$ have a finite correlation length $\xi_v$ (which is obviously fulfilled in the liquid range due to the high structural disorder), one can for $k\rightarrow 0$ make the replacement $\exp(\ldots)\rightarrow 1$, because for pairs $(j,j')$ such that $|\bm r_j-\bm r_{j'}|\gg \xi_v$, the actual signal is negligibly small because of the decoupled product $A_jA_{j'}$ of zero-mean variables. Therefore, one can safely write that
\begin{align}
  S_v(k)&\sur{\simeq}{k\rightarrow 0} \frac{k^2N^{-1}}{3}\left\lan\left(\sum_j\bm \tau_j\right)^2\right\ran=\frac{k^2N^{-1}}{3}\lan\bm P^2\ran
\end{align}
(where isotropy of space has been invoked). Therefore, the fact that $S_v(k)$ is $\propto k^4$ is directly related to the vanishing of $N^{-1}\lan \bm P^2\ran$ in the thermodynamic limit $N,V\rightarrow \infty$ with $V/N=v=\rm C^t$. We show in the next appendix that for a finite system with a external boundary $\sim V^{2/3}$, one should expect $\lan \bm P^2\ran\sim N^{2/3}$ from the bulk conservation of $\bm P$, and thence the asymptotic proof of our result.

\bigskip

It is interesting to consider also the $k^2$ behaviour of $S_i(k)$. First, one can notice that it is easily demonstrated using the Cauchy-Schwartz inequality $|S_i(k)|\leq \sqrt{S_{v}(k)S(k)}$, provided one assumes a regular behaviour of $S_i(k)$ near $k=0$. If however one tries to follow a more explicit route, one faces the fact that $S_i(k)$ for small $k$ has something to do with the large wavelength density fluctuations (which prevents for instance any small $k$ expansion of the exponential terms). To see that, we write
\begin{align}
  S_i(k)&\sur{\simeq}{k\rightarrow 0}-i\bm k\cdot \left\lan \bm \tau_j\sum_{j'}  e^{i\bm k \cdot \bm r_{j'j}}\right\ran\\
&=k^2\left\lan \sum_{j'}(\bm \tau_j\cdot\bm r_{j'j})\frac{\sinc(kr_{jj'})-\cos(kr_{jj'})}{(kr_{jj'})^2}\right\ran\label{c20}
\end{align}
where $\bm r_{j'j}=\bm r_j-\bm r_{j'}$ and the second line is obtained via the isotropy of space. It is tempting to take the $k=0$ value inside the average, and obtain a value $\frac 13\lan \sum_{j'}\bm\tau_j\cdot\bm r_{j'j}\ran$ for the leading $k^2$ coefficient. But this is not correct, because this term is ill-defined as regards the convergence. To avoid this divergence, one introduces the conditional probability $P(\Ga|\bm\tau_j)$ over the entire phase space, $\Ga=(\bm r_i)_{(i=1,N)}$, and writes \myref{c20} like
\begin{align}
  S_i(k)&\sur{\simeq}{k\rightarrow 0}k^2v^{-1}\int d^3 r\frac{\sinc(kr)-\cos(kr)}{(kr)^2}\bm r\cdot\left\lan\bm \tau_j g_{\bm \tau_j}(\bm r)\right\ran_{\bm\tau_j}\\
g_{\bm \tau_j}(\bm r)&=v\int d\Ga P(\Ga|\bm\tau_j)\sum_{i\neq j}\de(\bm r-(\bm r_i-\bm r_j))
\end{align}
where  $\lan\cdot\ran_{\bm \tau_j}$ means a average over the different values of $\bm \tau_j$ (with the associated equilibrium  probability). One has on the one hand $\lan\bm\tau_j\ran_{\bm\tau_j}=\lan\bm\tau_j\ran=\bm 0$, and $g_{\bm \tau_j}(\bm r)\rightarrow 1$ for large $|\bm r|$ on the other hand. On assuming as always a finite correlation length into the fluid, one can write
\begin{align}
  S_i(k)&\sur{\simeq}{k\rightarrow 0} \frac{k^2v^{-1}}{3}\int d^3r \bm r\cdot \lan \bm\tau_j [g_{\bm \tau_j}(\bm r)-1]\ran_{\bm \tau_j},\label{coeffk2}
\end{align}
an expression which is now well-defined with a convergent integral.
Qualitatively, one understands the fact that the $k^2$ coefficient is positive noticing that given a value of $\bm\tau_j$, the  particles surrounding $j$ and close to the axis parallel to $\bm\tau_j$ are typically ``close'' to $j$ if one follows the tail of the vector, and ``far'' from $j$ is one follows the opposite direction. Finally, let us remark that a similar expression as \myref{coeffk2} for the $k^4$ coefficient of $S_v(k)$ could be written, but would be of few interest, for correlators build up with conditional probabilities are awkward to compute and interpret.

\subsection{Appendix : Conservation of the geometrical polarisation}\label{App_pol_2}

(We deal here with 3D systems, but the generalization to $n\geq 2$ dimensions is obvious).
For an infinite system, we will show that $\bm P$, formally defined by Eq. \myref{Pdef}, is conserved. To this end, we use the formula
\begin{align}
  \bm \nab_j U_i&=-\frac{S_{ij}}{r_{ij}}\bm s_{ji}\label{gradU}
\end{align}
 demonstrated in appendix \ref{AppMakse}, which is appropriate only if (i) one has $i\neq j$, and (ii) the Voronoi cells of particles $i$ and $j$ share a common facet (otherwise $\bm \nab_j U_i=\bm 0$). In this formula, $S_{ij}$ is the area of the common facet (notice that this symbol will be used to term both the facet and its area; if this facet does not exist, $S_{ij}=0$ will be assumed whenever necessary), $r_{ij}=|\bm r_{ij}|=|\bm r_j-\bm r_i|$ and $\bm s_{ji}$ is the vector starting from the particle $j$ and pointing to the centroid of the common facet $S_{ij}$. If $i=j$, one can use the global conservation of the volume  and write
 \begin{align}
   \bm\nab_jU_j&=-\sum_{i\neq j}\bm \nab_jU_i
 \end{align}
Therefore, one has
 \begin{align}
   \pa_{z_j}\sum_i U_i\bm r_i&=U_j\bm e_z-\sum_{i\neq j}\bm r_{ji}\frac{S_{ij}}{r_{ij}}(\bm s_{ji}\cdot \bm e_z)
 \end{align}
Now, for each facet of the cell $j$, with $\bm r$ a running vector joining the particle $j$ to a point of $S_{ij}$, we have
\begin{align}
  \int_{S_{ij}}d\bm S (\bm r\cdot&\bm e_z)=\nonumber\\
&\left\{\begin{array}{l}\displaystyle\frac{\bm r_{ji}}{r_{ji}}\int_{S_{ij}}d\bm S (\bm r\cdot\bm e_z)=\frac{\bm r_{ji}}{r_{ji}}S_{ij}(\bm s_{ji}\cdot\bm e_z)\\
\displaystyle\int_{P_{ji}} d^3r \bm \nab(\bm r\cdot\bm e_z)-\int_{(\pa P_{ji})\setminus S_{ij}}d\bm S (\bm r\cdot\bm e_z)
  \end{array}\right.\label{2f}
\end{align}
The second relation is also 
\begin{align}
P_{ji}\bm e_z-\int_{(\pa P_{ji})\setminus S_{ij}}d\bm S (\bm r\cdot\bm e_z)\label{rajoutee}
\end{align}
where $P_{ji}$ is the (volume of) the pyramid with the particle $j$ at its summit and with $S_{ij}$ for base. The second relation comes from $\displaystyle\int_V d^3\tau \bm \nab \phi=\oint_{\pa V} \phi d\bm S$, a corollary of the Green-Ostrogradsky theorem. As soon as it is realized that $U_j$ is constructed with the different pyramids $(P_{ji})_i$, a lateral edge of a pyramid being actually a lateral edge of exactly one another pyramid, the last term of  \myref{rajoutee} cancel upon summation over $i$, and we get from  \myref{2f} that
\begin{align}
  \sum_{i}\frac{\bm r_{ji}}{r_{ji}}S_{ij}(\bm s_{ji}\cdot\bm e_z)=\sum_i{P_{ji}}\bm e_z=U_j\bm e_z
\end{align}
which yields $\pa_{z_j}\sum_i U_i\bm r_i=\bm 0$. As $j$ is arbitrary, as well as the choice of the $z$ coordinate, one concludes that the vector $\sum_iU_i\bm r_i$ is a constant. Besides, the vector $\sum_{i}U_i\bm s_i$ is nothing but the entire volume times a vector pointing to its geometrical center, therefore a constant, whence one concludes that $\bm P$ itself is conserved. The isotropy of space imposes moreover $\bm P=\bm 0$.

If the system is finite and large ($N\gg 1$), with ``normal'' boundaries scaling like $N^{2/3}$, $\bm P$ is not necessarily conserved, due to the fact that the pyramidal structure of Voronoi cells is no longer fulfilled for the boundary cells. Therefore, one expects fluctuations of $\bm P$ driven by $\sim N^{2/3}$ boundary cells. If one assumes rapid decorrelations among the boundary polarisation vector $\tau_i$, one should have $\lan\bm P^2\ran\sim N^{2/3}$, i.e. very weak fluctuations, and a rapid asymptotic behaviour for growing $N$ (see preceding section).

\bigskip

Finally, the preceding demonstration highlights some useful, non obvious equalities. Whatever the vector $\bm v$, one has
\begin{align}
  U_j\bm v&=\sum_{i\neq j}\hat{\bm r}_{ji}S_{ij}(\bm s_{ji}\cdot\bm v)=\sum_{i\neq j}(\hat{\bm r}_{ji}\cdot\bm v)S_{ij}\bm s_{ji}\label{coroutile}
\end{align}
(where $\hat{\bm r}_{ji}=\bm r_{ji}/r_{ji}$). A corollary of this is 
\begin{align}
  \forall\ (\bm v,\bm w),\ \ \bm v\cdot\bm w=0\Rightarrow \sum_{i\neq j}(\hat{\bm r}_{ji}\cdot\bm v)S_{ij}(\bm s_{ji}\cdot\bm w)=0
\end{align}

\bibliography{toutvenant,voronoi,localbib}
\end{document}